\documentclass[%
 aip,
 apl,
 amsmath,amssymb,
 reprint,%
]{revtex4-1}

\usepackage{graphicx}
\usepackage{dcolumn}
\usepackage{siunitx}
\usepackage{bm}

\usepackage{hyperref}
\hypersetup{colorlinks,linkcolor={blue},citecolor={blue},urlcolor={blue}}

\usepackage{mathptmx}
\usepackage{etoolbox}
\usepackage{scalerel}

\makeatletter
\def\@email#1#2{%
 \endgroup
 \patchcmd{\titleblock@produce}
  {\frontmatter@RRAPformat}
  {\frontmatter@RRAPformat{\produce@RRAP{*#1\href{mailto:#2}{#2}}}\frontmatter@RRAPformat}
  {}{}
}%
\makeatother

\begin{document}

\newcommand{\tr}{{\text{tr}}}
\newcommand{\inj}{{\text{inj}}}

\title{Over-Barrier Photoelectron emission with Rashba Spin-Orbit Coupling}

\author{Bi Hong Tiang}
\affiliation{Science, Mathematics and Technology (SMT) Cluster, Singapore University of Technology and Design (SUTD), Singapore 487372}%

\author{Yee Sin Ang}
\thanks{Authors to whom correspondence should be addressed: 
yeesin\_ang@sutd.edu.sg and ricky\_ang@sutd.edu.sg}
\affiliation{Science, Mathematics and Technology (SMT) Cluster, Singapore University of Technology and Design (SUTD), Singapore 487372}%

\author{L. K. Ang}
\thanks{Authors to whom correspondence should be addressed: 
yeesin\_ang@sutd.edu.sg and
ricky\_ang@sutd.edu.sg}
\affiliation{Science, Mathematics and Technology (SMT) Cluster, Singapore University of Technology and Design (SUTD), Singapore 487372}%

\begin{abstract}


We develop a theoretical model to calculate the quantum efficiency (QE) of photoelectron emission from materials with Rashba spin-orbit coupling (RSOC) effect. 
In the low temperature limit, an analytical scaling between QE and the RSOC strength is obtained as QE $\propto (\hbar\omega-W)^2+2E_R(\hbar \omega-W) -E_R^2/3$, where $\hbar\omega$, $W$ and $E_R$ are the incident photon energy, work function and the RSOC parameter respectively.  
Intriguingly, the RSOC effect substantially improves the QE for strong RSOC materials.
For example, the QE of Bi$_2$Se$_3$ and Bi/Si(111) increases, by 149\% and 122\%, respectively due to the presence of strong RSOC.
By fitting to the photoelectron emission characteristics, the analytical scaling law can be employed to extract the RSOC strength, thus offering a useful tool to characterize the RSOC effect in materials.
Importantly, when the traditional Fowler-Dubridge model is used, the extracted results may substantially deviate from the actual values by $\sim90\%$, thus highlighting the importance of employing our model to analyse the photoelectron emission especially for materials with strong RSOC.
These findings provide a theoretical foundation for the design of photoemitters using Rashba spintronic materials.



\end{abstract}

\maketitle




Photoemission \cite{Einstein} has significant technological applications in photoemission spectroscopy \cite{spectroscopy1,spectroscopy3}, x-ray generators \cite{xray}, and photodetectors \cite{photodetector1,photodetector2,photodetector3,photodetector4,photodetector5}. 
One key property for studying photoelectron emission from materials is the quantum efficiency (QE), defined as the ratio of the number of emitted photo-electrons to the number of incident photons \cite{zhou2021quantum}. 
The  prevailing method for determining QE in photoemission is based on the Fowler-Dubridge (FD) model \cite{fowler,dubridge}, which was developed in the early twentieth century based on the assumption of a bulk material with parabolic energy dispersion. 
Multiple theoretical revisions of FD model have been subsequently developed, including the impact of laser heating on QE \cite{zhou2021quantum}, and generalized thermal-field-photoemission model \cite{jensen2002generalized}. 
Particularly, the use of quantum tunneling model have greatly enriches the photoemission physics over a wide range operating conditions to include various effects such as non-equilibrium heating at sub-ps time scale \cite{PhysRevB.78.224112}, few cycles \cite{PhysRevB.86.045423,photomodel3}, consistent time-dependent tunneling model \cite{PhysRevB.88.195434,zhou2020quantum,photomodel1}, dielectric coating \cite{acsnano.0c03406,zhou2022theory}, and dual-lasers \cite{photomodel2}.

These photoelectron emission models were however limited traditional metallic bulk materials that emerging nanomaterials, such as topological insulators \cite{topo1,topo2}, graphene \cite{photopgrahene1,photographene2} and spintronic materials, are relatively studied if the FD model is valid.
Unlike photoelectron emission, new models \cite{inf2.12168} for thermionic emission \cite{PhysRevApplied.3.014002,PhysRevLett.121.056802,PhysRevApplied.12.014057} and field emission \cite{PhysRevB.104.245420,PhysRevApplied.16.064025,10.1063/5.0137400} models for novel materials have been developed.


Among the emerging materials, Rashba spintronic materials (RSM) have garnered significant attention due to their unique electronic properties, such as tunability \cite{tunable1,tunable2,tunable3,tunable4} and electronic band-splitting without an external magnetic field \cite{rashba}.  RSM belong to a class of systems characterized by strong Rashba spin-orbit coupling \cite{manchon2015new,bihlmayer2022rashba},  which was first proposed in 1959 \cite{rashba1960properties}. The Rashba effect results from the combination of spin-orbit interaction and breaking of inversion symmetry induced by an external electric field in a direction orthogonal to the two-dimensional material's surface. The Rashba effect has been studied and experimentally observed in various materials, such as topological insulator Bi$_2$Se$_3$ \cite{tunable3}, metal \cite{aumassandrashba,Ir,metalrashba3}, graphene \cite{graphenerashba,graphenegiantrashba}, InAlAs/InGaAs \cite{InAlAs}, BiTeI\cite{ishizaka2011giant} and perovskite \cite{rashbaperovskites,perovskite2}. 
The unique electronic structure of RSM can be harnessed to create spin-polarized currents \cite{Rashbacoupling,spintronics1}, control the spin polarization and the direction of the spin currents \cite{spintronics2}. Thus, RSM are promising candidates for the development of spintronic devices, such as spin transistors\cite{spintransistor} and spin-based memory devices \cite{spinmemory}. 

Photoemission spectroscopy offers a valuable method for studying RSM. Specifically, the angle-resolved photoemission spectroscopy (ARPES) and the spin-resolved photoemission spectroscopy (SRPES) can be used to measure the Rashba splitting \cite{yaji2010large,spectroscopy2,spectroscopy4,spectroscopy5,spectroscopy6,spectroscopy7},  to examine the spin-polarized photoemission  \cite{spectroscopy6,spectroscopy7} and to investigate the electronic band structures of RSM \cite{srpes1,srpes2}. 
Consistent theoretical model for photoelectron emission for RSM remains limited and mostly bounded by the classic FD model, which does not include the spins splitting effects in the formulation. 
For instance, ARPES and SRPES utilize the assumption of parabolic electron dispersion to estimate the Rashba parameter $\alpha_R$ based on the values of measured momentum offset $k_0=\alpha_R m/\hbar^2$ and Rashba energy $E_R=\hbar^2 k_0^2/2m$. 
Such approach has, however, overlooked the spin splitting within the dispersion and may introduce errors in the estimation of the Rashba parameter \cite{ishizaka2011giant}. 


In this work, we develop a theoretical model to study how the Rashba spin-orbit coupling (RSOC) will modify the quantum efficiency (QE) of photoelectron emission from Rashba spintronic materials (RSM). In low temperature limit, our model presents an analytical scaling of QE $\propto (\hbar\omega-W)^2+2E_R(\hbar \omega-W) -E_R^2/3$, where $\hbar\omega$, $W$ and $E_R$ are the incident photon energy, work function and the RSOC parameter respectively. 
This finding suggest that the traditional Fowler-Dubridge (FD) model is no longer valid for photoelectron emission for RSM.
 Our model reveals that RSOC substantially increase the QE for materials with a strong RSOC strength and photon energy close to an effective work function ($W+E_R$). 
 For instance, QE of Bi$_2$Se$_3$ increases by up to 95\% due to the presence of strong RSOC. 
 Substantial error appears if the traditional FD model is used to characterize the photoelectron emission from RSM in order to estimate its RSOC strength.

\begin{figure}[t!]
\centering
  \includegraphics[width=0.45\textwidth]{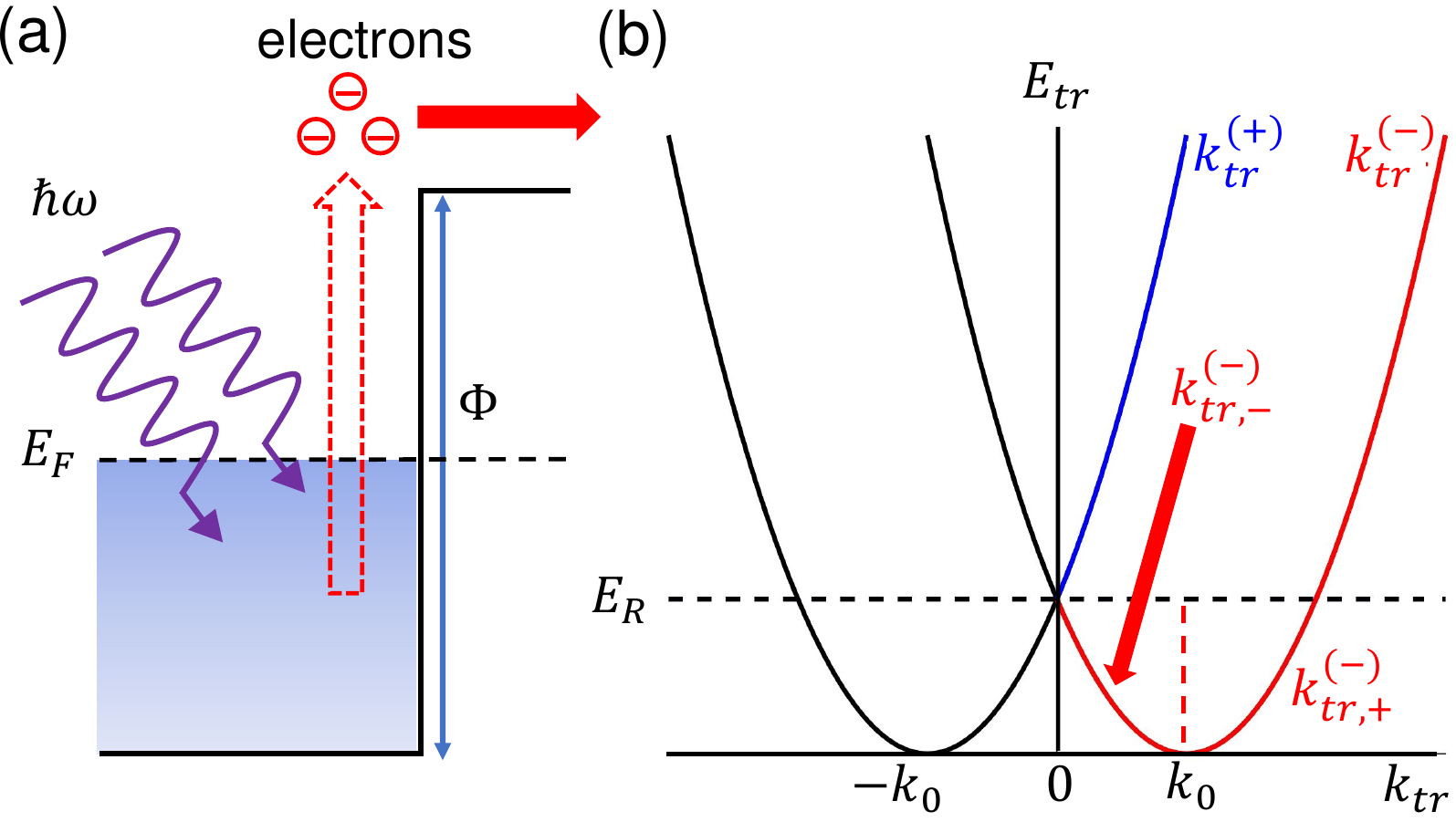}
    \caption{\label{fig:model, dispersion} Schematic diagrams of (a) the over-barrier photoemission; and (b) the electron energy dispersion of a Rashba spintronic materials plotted with $k_\text{inj} = 0$. The notation $\pm$ corresponds to the electronic band $s=\pm 1$. }
\end{figure}

First, the QE of photoemission in the FD model is \cite{zhou2020quantum}
\begin{equation}\label{eq: QE metal}
\text{QE}=a_1(1-r) A_1T^2 F\left(\frac{\hbar\omega-W} {k_BT}\right),
\end{equation}
where the QE is a function of incident photon energy $\hbar \omega$, material's work function $W$ and temperature $T$. 
Refer to Eq.~(\ref{eq: QE general}) for the detailed explanation of Eq.~(\ref{eq: QE metal}) and the defined parameters. 
In our derivation, we employ the FD approach to derive its QE, but replace the standard parabolic electron dispersion with the proper dispersion of RSM as shown in Eq.~(\ref{eq:rashbadispersion}). 
Assuming the incident photon is in $z-$direction and the electrons obey Fermi-Dirac distribution, the QE of photoelectron emission is calculated by
\begin{align}\label{eq: heaviside formalism}
    \text{QE} &\propto \frac{g}{(2\pi)^3}  \int d\phi \int dk_{\text{tr}}\; k_{\text{tr}}\nonumber\\
    &\qquad \; \int dk_z \;  \dfrac{\Theta \big[E_{\inj} -(E_F+W-\hbar \omega)\big]}{\exp{\big[(E_{\text{tr}}+E_{\text{inj}}-E_F)/k_BT\big]}+1}  
\end{align}
where $g$ is the degeneracy of electron's states, $E_F$ is the Fermi energy, $W$ is work function, $\hbar \omega$ is the injected photon energy, $k_B$ is the Boltzmann constant, $\Theta \big[E_{\inj} -(E_F+W-\hbar \omega)\big]$ is the Heaviside electron tunneling probability as the amount of energy required to overcome the barrier is reduced by the amount of injected photon energy [see Fig.~\ref{fig:model, dispersion}(a)].  
Based on the Rashba's system dispersion \cite{rashba}, we have
\begin{equation}\label{eq:rashbadispersion}
        E_{k,s}=\underbrace{\frac{\hbar^2}{2m} \bigg[(k_{\tr}+s k_0)^2\bigg]}_{ \scaleto{E_{\tr}}{8pt}}+\underbrace{\bigg[\frac{\hbar^2}{2m} k_z^2\bigg]}_{\scaleto{E_{\inj}}{10pt}}
\end{equation}
where $E_{k,s}$ is the total energy, $\displaystyle k_0=\alpha_R m/\hbar^2$ is the material-dependent parameter associated with the RSOC strength $\alpha_R$, $m$ is the electron effective mass $m$ and $\hbar$ reduced Planck's constant.
Here, $E_{tr}$ and $E_{\inj}$ is the energy in $xy$-plane and $z$-direction, respectively, $s=\pm 1$ corresponds to different electronic conduction band, $\Vec{k}=(k_x,k_y,k_z)$ is the wave vector, and $\displaystyle k_{\tr}=\sqrt{k_x^2+k_y^2}$.

To solve Eq.~(\ref{eq: heaviside formalism}), we determine the real solution of $k_{\tr}^{(s)}\geq 0$ in the Rashba dispersion as indicated in Fig.~\ref{fig:model, dispersion}(b). 
Let's define the Rashba energy as $\displaystyle E_R=\hbar^2 k_0^2/2m$. 
At $s=+1$, we only have  $k_{\text{tr}}^{(+)} = \sqrt{2mE_{\tr}}/\hbar -k_0$ for $ E_{\text{tr}} \geq E_R\;$. 
At $s=-1$, we obtain $k_{\text{tr},\eta}^{(-)} = \eta \sqrt{2mE_{\tr}}/\hbar +k_0 \;$ for $ E_{\text{tr}} \leq E_R$ where $\eta = \pm 1$, and $k_{\tr}^{(-)}=\sqrt{2mE_{\tr}}/\hbar +k_0$ for $E_{\text{tr}} \geq E_R$. 
Thus, Eq.~(\ref{eq: heaviside formalism}) becomes
\begin{widetext}
\begin{align}\label{eq: QE}
  \text{QE} \propto  \frac{1}{2\sqrt{2}\pi^2}  \frac{m^{3/2}}{\hbar^3} \Bigg\{  \int_{E_F+W-\hbar \omega}^{\infty} \frac{dE_{\text{inj}}}{\sqrt{E_{\inj}}} \; \int_{E_R}^{\infty} dE_{\text{tr}}\; f_{FD} 
    \quad +  \frac{\hbar  k_0}{\sqrt{2m}}\;\int_0^{E_R} \frac{dE_{\tr}}{\sqrt{E_{\tr}}} \;\int_{E_F+W-\hbar \omega}^{\infty} dE_{\text{inj}} \frac{f_{FD}}{\sqrt{E_{\inj}}} \Bigg\}   
\end{align}
where $\displaystyle f_{FD}=\dfrac{1}{\exp{\big[(E_{\text{tr}}+E_{\text{inj}}-E_F)/k_BT\big]}+1}$ denotes the Fermi-Dirac distribution. 
Note Equation.~(\ref{eq: QE}) can be rewritten as 
\begin{align}\label{eq: QE general}
\text{QE} =a_1(1-r)A_1 T^2\bigg[ F\bigg(\frac{\hbar\omega-W-E_R}{k_BT}\bigg)+2\sqrt{\frac{E_R}{k_BT}} \mathcal{R}(\hbar\omega,W,E_R,k_BT)\bigg]
\end{align},
where an experimental fitting parameter $a_1$ (like in the FD model) has been added. 
For example, $a_1 = 5 \times 10^{-18}\; \text{m}^2/ \text{A}$ for Cu \cite{zhou2020quantum}.
Here the parameter $r$ is the reflectivity which depends on the incident photon wavelength, angle, and refractive index of the materials.
The constant $A_1 = 120 \,\text{A}/(\text{cm}^2\text{K}^2)$ is the Richardson’s constant.
The two functions are the Fowler function $\displaystyle F(x)=\pi^2/6 +x^2/2-\exp{(-x)}+\exp{(-2x)}/2^2+\cdots$ for $x>0$ \cite{fowler,zhou2020quantum}, and a new normalized function of
\begin{eqnarray}
    \mathcal{R}(\hbar\omega,W,E_R,k_B T) = \int_0^{\sqrt{E_R/k_BT}} du \ln{\bigg[ 1+\exp{\left(\frac{\hbar \omega-W}{k_BT}-u^2\right)} \bigg] }.
\end{eqnarray}
We recover Eq.~(\ref{eq: QE metal}) from Eq.~(\ref{eq: QE general}) at $E_R$= 0 and thus converges to the FD model without RSOC for consistency purpose.

The RSOC effect or finite value of $E_R$ has two effects. 
Firstly, it raises the effective work function of the first term 
in Eq.(\ref{eq: QE general}) from $W$ to $W+E_R$.
The second term in Eq.~(\ref{eq: QE general}) serves the enhancement of QE due to finite $E_R$. 
Thus, the net effect of RSOC will depend on the laser wavelength, material's work function, temperature and RSOC strength as shown in the figures below. 
At low temperature limit $k_BT \ll \hbar\omega-W-E_R$ [see Supplementary materials (SM) for the complete derivation], Eq.~(\ref{eq: QE general}) can be further simplified to an analytical form of
\begin{align}\label{eq: QE zero T}
    \text{QE} (T \to 0)=\Theta(\hbar\omega-W-E_R)\;\frac{a_1(1-r) A_1 }{2 k_B^2} \;
     \bigg[(\hbar\omega-W)^2+2E_R(\hbar \omega-W) -E_R^2/3\bigg],
\end{align}
where $\Theta(x)$ is the Heaviside function.  From Fig.~\ref{fig: parameter plot and QE}(f), Eq.~(\ref{eq: QE zero T}) agrees well with Eq.~(\ref{eq: QE general}) at room temperature $T$ = 300 K and wavelength $\lambda$ = 200 nm as the error $E$ is less than 1 \%. However, when the wavelength or temperature increases, the error $E$ also increases. Thus, Eq.~(\ref{eq: QE zero T}) is a good approximation only when the ratio $\displaystyle \bigg(\dfrac{\hbar\omega-W-E_R}{k_B T}\bigg)$ is large. Importantly, the experimental fitting of Eq.~(\ref{eq: QE zero T}) allows for the determination of the work function $W$ and parameter $E_R$, which play a crucial role in characterizing the fundamental features of Rashba spintronic materials.

\begin{figure*}[t!]
\centering
  \includegraphics[width=\textwidth]{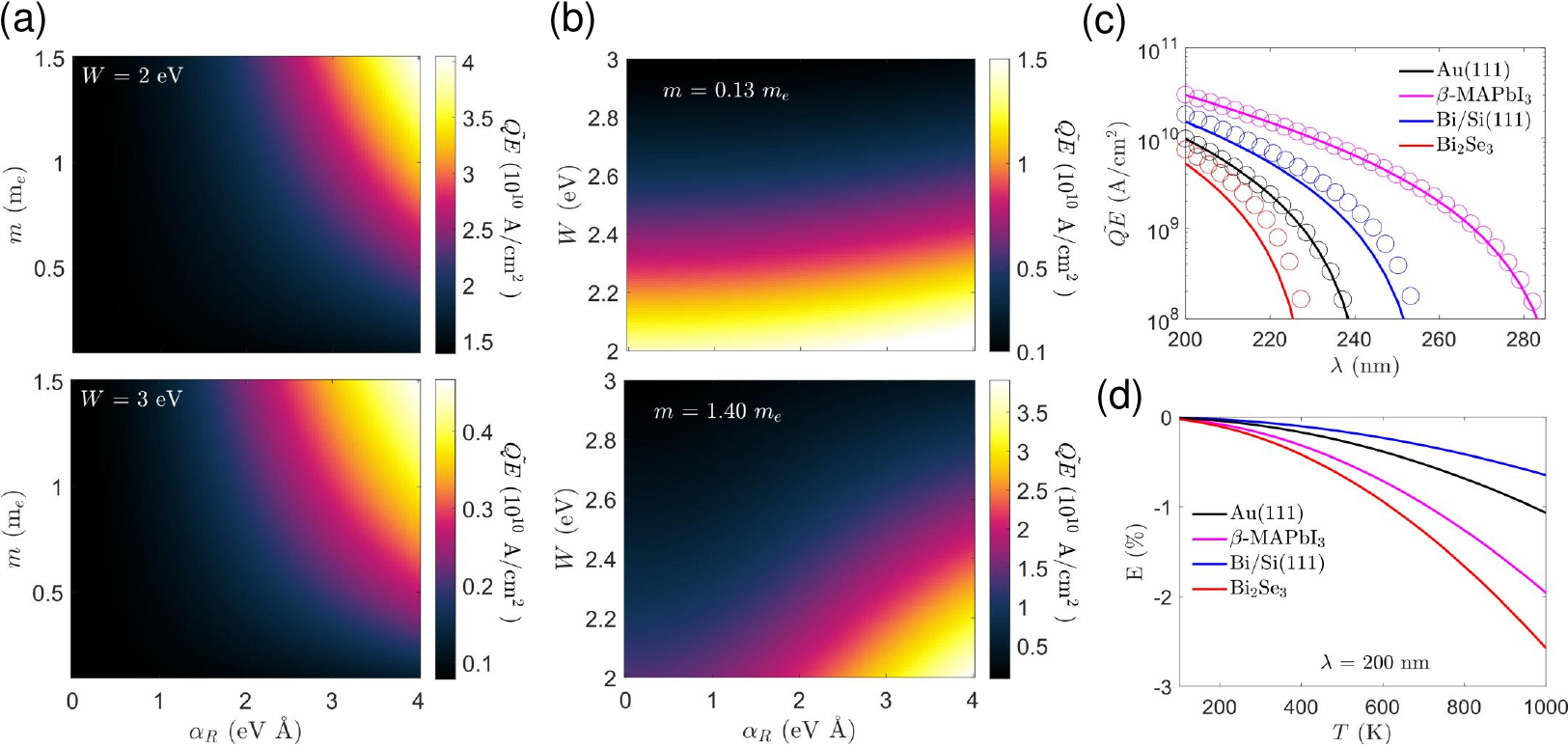}
\caption{\label{fig: parameter plot and QE} QE enhancement due to Rashba spin-orbit coupling effect (RSOC) at temperature $T$ = 300 K. (a) The $\tilde{QE} \equiv QE/a_1(1-r)$ as a function of electron effective mass ($m$) and RSOC strength ($\alpha_R$) is calculated for work functions $W = 2$ eV (top) and $W = 3$ eV (bottom). (b) The $\tilde{QE}$ as a function of work function ($W$) and RSOC strength ($\alpha_R$) is calculated for electron effective masses $m = 0.13 \;m_e$ (top); and $m = 1.40\; m_e$ (bottom). 
The incident light wavelength is set to a representative value of $\lambda$ = 375 nm. (c) The $\tilde{QE}$ as a function of laser wavelength $\lambda$ for different materials: $\tilde{QE}$ without RSOC (solid line) and $\tilde{QE}$ with RSOC (circle). (d) $E=[\text{QE}(T \to 0)-\text{QE}]/\text{QE}$ as a function of temperature ($T$), where $E$ is the relative error percentage between the numerical solution of $\tilde{QE}$ at $T$ up to 1000 K and the analytical $\tilde{QE}$ based on $T$=0 approximation. }
\end{figure*}
\end{widetext}
\begin{figure*}[t!]
\centering
  \includegraphics[width=\textwidth]{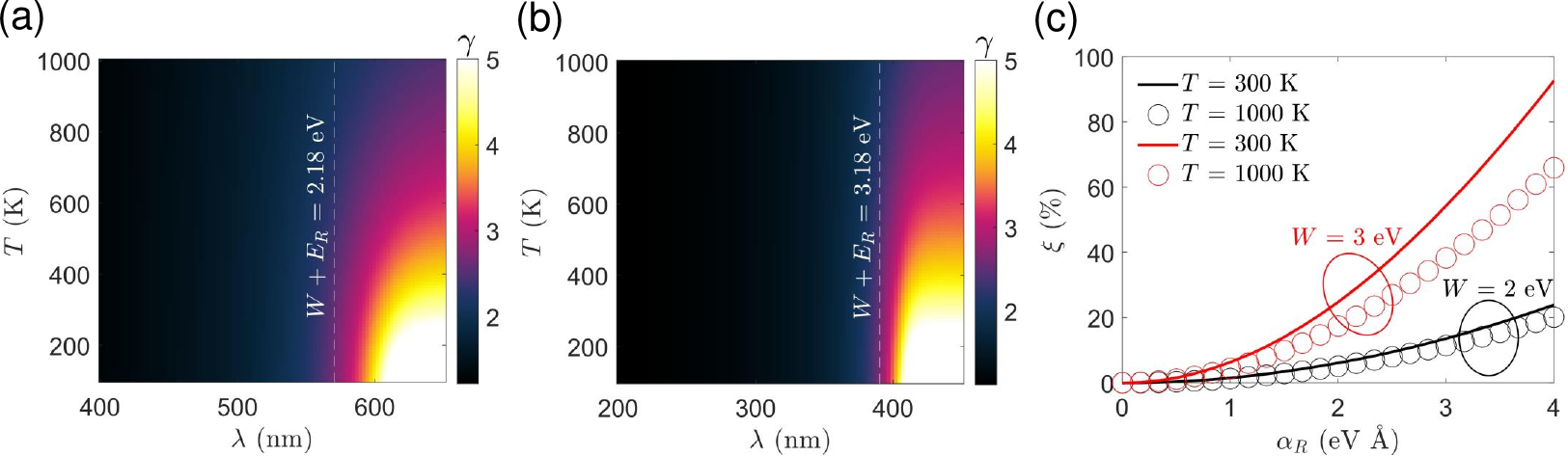}
    \caption{\label{fig: qe ratio} The effect of work function and incident laser wavelength in photoemission enhancement due to Rashba spin-orbit coupling (RSOC) effect. Here, we employ the values of electron effective mass and RSOC strength of Bi$_2$Se$_3$. $\gamma$ is the the enhancement due to RSOC as a function of laser wavelength ($\lambda$) and temperature ($T$) for work function (a) $W$ = 2 eV;  and (b) $W$ = 3 eV. The white dashed line indicates the incident wavelength corresponds to the threshold $W+E_R$. (c) The relative error $\xi$ as a function of RSOC strength ($\alpha_R$) at different temperature $T$= 300 K (solid line), 1000 K (circle) and laser wavelength $\lambda$ = 375 nm for work function $W$ = 2 eV (black) and $W$ = 3 eV (red).}
\end{figure*}

\begin{table}[t!]
\caption{\label{tab:physical constants}
Physical properties of selected materials.}
\begin{ruledtabular}
\begin{tabular}{ldddd}
Materials  & \multicolumn{1}{c}{$W$(eV)} 
  & \multicolumn{1}{c}{$E_R$ (meV)} & \multicolumn{1}{c}{$\alpha_R$ (eV \AA)} & \multicolumn{1}{c}{$m$ ($m_e$)}\\
\hline
Au (111) & 5.1 \cite{zhou2020quantum} & 2.1\cite{ishizaka2011giant}  & 0.33 \cite{ishizaka2011giant} & 0.29\\
$\beta$-MAPbI$_3$ &  4.28\cite{perovskiteworkfunction} & 12\cite{perovskiteRashba} & 1.5\cite{perovskiteRashba} & 0.10\\
Bi/Si(111) & 4.83\cite{siliconworkfunction} & 140 \cite{gierz2009silicon}& 1.37\cite{gierz2009silicon} & 1.14\\
Bi$_2$Se$_3$ & 5.4 \cite{bi2se3workfunction} & 180 \cite{RashbaBi2Se3}  & 4.0\cite{manchon2015new} & 0.17\\
\end{tabular}
\end{ruledtabular}
\end{table}

 The FD model suggests a dependence of QE $\propto (\hbar \omega -W)^2$ which agrees well with experimental data \cite{inf2.12168}. 
 Here, Eq.~(\ref{eq: QE zero T}) generalizes the QE of a photoemitter to the case of Rashba spintronic system, yielding an unconventional scaling of
 \begin{equation}
     \text{QE} \propto (\hbar\omega-W)^2+2E_R(\hbar \omega-W) -E_R^2/3.
 \end{equation}

To analyze the parametric dependence of photoemission QE in RSM, we numerically compute the QE for different set of physical parameters, including electron effective mass, work function, laser wavelength, temperature and RSOC strength [see Fig.~(\ref{fig: parameter plot and QE}) and Fig.~(\ref{fig: qe ratio})]. In Fig.~\ref{fig: parameter plot and QE}, we plot $\tilde{QE} \equiv QE/a_1(1-r)$ as $a_1$ and $r$ are some material-dependent parameters, which may be determined experimentally. 
For both low and high work function case [see Fig.\ref{fig: parameter plot and QE}(a)], we observe that increasing the electron effective mass $m$ or RSOC strength $\alpha_R$ will lead to an increase in QE. 
Figure~\ref{fig: parameter plot and QE}(b) show that a higher electron effective mass ($m$) will imply a more pronounced enhancement of $\tilde{QE}$, owing to the $m$-dependence of $E_R$. 
Since both the electron effective mass $m$ and RSOC strength $\alpha_R$ are positively correlated with $E_R$, increasing $E_R$ will lead to an increase in $\tilde{QE}$.  
Moreover, as depicted in Fig.~\ref{fig: parameter plot and QE}(a,b), $\tilde{QE}$ decreases with higher work function $W$ as expected as fewer electrons possess adequate energy to overcome the higher $W$.


In Fig.~\ref{fig: parameter plot and QE}(c), we specifically pick four materials with different $E_R$ as shown in Table.~(\ref{tab:physical constants}):  Au(111), $\beta$-MAPbI$_3$, Bi/Si(111) and Bi$_2$Se$_3$.
The calculated results in the figure compare the calculations with (symbols) and without the RSOC effect.
The comparison suggests significant difference for Bi/Si(111) and Bi$_2$Se$_3$ due to larger values of $E_R$ for both materials.
The figure also shows that the $\tilde{QE}$ improvement is more significant at longer wavelengths. 

To study the wavelength dependence of QE improvement, we compute the ratio of QE with RSOC to the QE without RSOC ($\gamma$), i.e. 
%
%
\begin{equation} \label{eq:xi}
   \gamma =\dfrac{ F\bigg(\dfrac{\hbar\omega-W-E_R}{k_BT} \bigg)+2\sqrt{\dfrac{E_R}{k_BT}}\mathcal{R}(\hbar\omega,W,E_R,k_BT)} {F\bigg(\dfrac{\hbar\omega-W}{k_BT}\bigg)}
\end{equation}
%
%
\noindent
where the functions $F$, $\mathcal{R}$ and parameter $B_1$ are defined in Eq.~(\ref{eq: QE general}). For over-barrier photoemission $\hbar \omega \geq W+E_R$, Fig.~\ref{fig: qe ratio} (a,b) reveals that the presence of strong Rashba spin-orbit coupling (RSOC) results in a QE enhancement ranging from approximately 15\% to 150\%. 
The QE enhancement rises at longer laser wavelength, which is consistent with Fig.~\ref{fig: parameter plot and QE}(c). 
For laser energies lower than the threshold of $W+E_R$, Fig.~\ref{fig: qe ratio} (a,b) predicts a huge QE enhancement, estimated to be at least $\sim150\%$. 
There is also a significant decrease in the QE enhancement as the temperature increases, suggesting that thermal excitation has emerged as a crucial factor affecting QE enhancement. Hence, in both low and high photon energy scenarios, RSOC presents a compelling opportunity to boost the QE of electron emission.

Finally, for photoemission experiment, the value of the prefactor $a_1$ is fitted by measuring QE. 
Here, we estimate the relative error percentage $\xi$ for the values of $a_1$ extracted via the traditional FD model ($a_1^{(FD)}$) and our model ($a_1^{(R)}$), in which the $\xi$ is given by 
\begin{equation}\label{eq: relative error}
    \xi = \dfrac{a_1^{(FD)}-a_1^{(R)}}{a_1^{(R)}} =\gamma-1
\end{equation}
%
%
%
%
where $\gamma$ is  defined in Eq.~(\ref{eq:xi}). 
Figure \ref{fig: qe ratio}(c) shows that $\xi$ will increase with RSOC strength.
The error is also more significant at lower temperature and higher work function.
As an example, the peak $\xi$ = 90$\%$ occurs at work function $W$ = 3 eV, $T$ = 300 K and RSOC strength $\alpha_R$ = 4 eV \AA. 
This discrepancy in extracted values of the $a_1$ parameter between our model and Fowler-Dubridge model is significant, and hence emphasizing the importance of using the appropriate model when examining photoemission measurement data. Furthermore, when the laser energy approaches the threshold of $W+E_R$, an increase in temperature will cause in a notable decrease in the values of $\xi$ as depicted in Fig.~\ref{fig: qe ratio}(c), which suggests the impact of RSOC on QE diminishing at higher temperatures . 
Thus, our results predicts a substantial QE improvement at low temperature (room temperature) for materials with a strong RSOC strength and for incident photon energy near the threshold $W+E_R$. 
Conversely, materials with a weak RSOC strength or high incident photon energy exceeding the $W+E_R$ threshold exhibit a low QE enhancement. 
Our model can facilitate the design of high-performance photodetectors by tuning the RSOC strength \cite{tunable1,tunable2,tunable3,tunable4} to advance both fundamental knowledge and applications. 



In summary, we have developed a theoretical model to calculate the quantum efficiency (QE) of photoelectron emission from Rashba spintronic materials with RSOC effect.  
In low temperature limit [see Eq.~(\ref{eq: QE zero T})], our model presents an analytical expression of QE $\propto (\hbar\omega-W)^2+2E_R(\hbar \omega-W) -E_R^2/3$, where $\hbar\omega$, $W$ and $E_R$ are the incident photon energy, work function and the RSOC parameter respectively. 
This unique scaling law can be conveniently employed to gauge the RSOC strength in RSM, thus providing a useful tool for characterizing Rashba spintronic materials. 
Importantly, RSOC substantially improves the QE for materials with a strong RSOC strength and for photon energy close to an effective work function ($W+E_R$) [see Fig.~\ref{fig: qe ratio}(a,b)]. 
For instance, QE of Bi$_2$Se$_3$ increases by up to 149\% and QE of Bi/Si(111) increases by up to 122\% , due to the presence of strong RSOC. 

It is essential to employ our proposed model here to examine photoelectron emission data as the prefactor $a_1$ parameter from the traditional Fowler-Dubridge (FD) model can substantially deviate from the actual values up to 90\% [see Fig.~\ref{fig: qe ratio}(c)].  
These findings pave the way for the advancement and characterization of photodetectors that rely on Rashba spintronic materials.
For future works, the formulation in this work is limited to a specific range of incident photon wavelengths, where the photon energy is greater than the work function of the materials, where the single-photon absorption dominates the photoemission process \cite{zhou2020quantum}. 
The model can be expanded to include multiphoton absorption and optical tunneling regimes by considering longer incident photon wavelengths. 


See supplementary material for the complete derivation of Eq.~(\ref{eq: QE zero T}).

\section*{Acknowledgements}
This work is supported by the Singapore A*STAR IRG grant (A2083c0057). Y. S. A. is supported by the Singapore Ministry of Education Academic Research Fund Tier 2 (Award No. MOE-T2EP50221-0019)


\section*{AUTHOR DECLARATIONS}
\section*{Conflict of Interest}
The authors have no conflicts to disclose.

\section*{Author Contributions}
 \textbf{Bi Hong Tiang}: Methodology (equal), Investigation (lead), Writing - Orignial Draft (lead).
  \textbf{Lay Kee Ang}: Conceptualization (lead), Supervision (equal), Writing - Review \& Editing (equal).
 \textbf{Yee Sin Ang}: Methodology (equal), Supervision (lead) , 
 Writing - Review \& Editing (lead).

\section*{DATA AVAILABILITY}
The data that support the findings of this study are available
from the corresponding author upon reasonable request.


%

\end{document}